\documentclass[12pt,preprint]{aastex}

\slugcomment{appeared in ApJL v636 (Jan. 10, 2006)}

\shorttitle{The Effects of Metallicity and Grain Size on GI's}
\shortauthors{Cai et al.}

\begin{document}

\title{The Effects of Metallicity and Grain Size \\
  on Gravitational Instabilities in Protoplanetary Disks}

\author{Kai Cai, Richard H. Durisen, 
Scott Michael, Aaron C. Boley}
\affil{Astronomy Department, Indiana University,
    Bloomington, IN 47405}
\email{kai@astro.indiana.edu}

\notetoeditor{Email addresses for other authors from IU are not provided 
due to limited space. If you want, they're: 
durisen@astro.indiana.edu; 
scamicha@astro.indiana.edu (Scott Michael), acboley@astro.indiana.edu 
(Aaron Boley)}

\author{Annie C. Mej\'{\i}a}
\affil{Department of Astronomy, University of Washington, Box 351580, 
Seattle, WA 98195-1580}
\email{acmejia@astro.washington.edu}

\author{Megan K. Pickett}
\affil{Department of Chemistry and Physics, Purdue University Calumet, 
2200 169th St., Hammond, IN 46323}
\email{pickett@astro.calumet.purdue.edu}

\and

\author{Paola D'Alessio}
\affil{Centro de Radioastronomia y de Astrofisica, Apartado Postal 72-3, 
58089 Morelia, Michoacan, Mexico} 
\email{p.dalessio@astrosmo.unam.mx}

\begin{abstract}
Observational studies show that the probability of finding
gas giant planets around a star increases with the star's metallicity. 
Our latest simulations of disks undergoing gravitational instabilities (GI's) 
with realistic radiative cooling indicate that protoplanetary disks with
lower metallicity generally cool faster and thus show stronger overall 
GI-activity. More importantly, the global cooling times in our simulations 
are too long for disk fragmentation to occur, and the disks do not 
fragment into dense protoplanetary 
clumps. Our results suggest that direct gas giant planet 
formation via disk instabilities is unlikely to be the mechanism that 
produced most observed planets. Nevertheless, GI's may still play an 
important role in a hybrid scenario, compatible with the observed metallicity 
trend, where structure created by GI's accelerates planet 
formation by core accretion. 
\end{abstract}


\section{INTRODUCTION}

The past decade has seen the discovery of over 150 exoplanets 
(http://www.obspm.fr/planets).  One statistical trend that has emerged
from these data is that the probability of finding a gas giant planet around 
a star with current techniques increases with the host star's metallicity
\citep{santos01, fv05}. As shown by \citet{fv05}, 
the high metal content of planet host stars seems to be primordial. 
Therefore, this trend, if real \citep{sozzetti05}, indicates that 
short-period 
gas giant planets are more likely to occur in metal-rich than 
in metal-poor protoplanetary disks.
The two contending theories for gas giant formation are core accretion plus 
gas capture \citep{pollack96} and disk instability \citep{boss02, mayer04}.
Calculations show that the metallicity relation can be
explained within the framework of core accretion  
\citep[e.g.,][]{idalin04,kornet05}.  For disk instability, \citet{boss02} finds 
that, in his three-dimensional hydrodynamics disk
simulations with radiative cooling, clump formation by disk instability 
occurs for all metallicities over the range 0.1 to 10 Z$_{\odot}$, due to 
rapid cooling by convection \citep{boss04}, and he attributes the 
abundance of short period gas giants around high metallicity stars 
to migration \citep{boss05}, a mechanism also invoked by \citet{idalin04}
to explain part of the metallicity correlation. By contrast, \citet{mejiaphd}, 
who uses a somewhat more sophisticated treatment of radiative 
boundary conditions, finds much longer cooling times and no  
fragmentation into dense clumps in her disk instability simulations. 
Here we report results of 
new disk calculations based on Mej\'ia's methods in which the opacity is 
varied by using different metallicities and grain sizes. Even over a much
narrower range of metallicities than considered by \citet{boss02}, we find 
that the strength of the GI's does vary noticeably and that disk 
fragmentation is not seen for any metallicity or grain size tested.

\section{METHODS}

We conduct protoplanetary disk simulations
using the Indiana University Hydrodynamics Group code 
\citep[see][]{pickett98,pickett00,pickett03,mejiaphd,mejia05},
which solves the equations of hydrodynamics in conservative form
on a cylindrical grid $(r,\phi,z)$ to second order in space and time 
using an ideal gas equation of state.  Self-gravity and shock mediation by 
artificial bulk viscosity are included. Reflection
symmetry is assumed about the equatorial plane, and free outflow 
boundaries are used along the top, outer, and inner edges 
of the grid.

We adopt the treatment of radiative physics detailed in \citet{mejiaphd}
with few modifications. Let $\tau_R$ be the optical depth, defined by 
using the Rosseland mean opacity measured vertically down from above. 
Energy flow in cells with $\tau_R > 2/3$ is calculated in all three directions 
using flux-limited diffusion \citep{bodenheimer90}. Cells with 
$\tau_R <$ 2/3, in the disk atmosphere and in the outer disk, cool radiatively 
using an optically thin LTE emissivity. Atmosphere heating by high-altitude 
shocks and upward moving photons from the photosphere are  
included. In this paper, we also assume that an external 
envelope heated by the star \citep{natta93, dalessio97} shines vertically 
down on the disk.  This IR irradiation is characterized by a black body flux 
with a temperature $T_{irr}$. The optically thick and thin regions 
are coupled, over one or two cells, by an Eddington-like grey atmosphere 
fit that defines the boundary flux for the diffusion approximation. 
The opacities and molecular weights for a solar 
composition are from \citet{dalessio01}, with a power-law grain size 
distribution of $n(a) \sim a^{-3.5}$ ranging from 0.005 $\mu$m to a 
largest grain size $a_{max}$ that can be varied. To model variations in 
metallicity Z,  the mean opacities are multiplied by a factor 
$f_\kappa =$ Z/Z$_{\odot}$, as was done by \citet{boss02}. 
Tests of our radiative scheme for a vertically stratified gas layer  
with a constant gravity, a constant input flux at the base, and a grey
opacity show relaxation to an Eddington-like solution with the 
correct flux from the photospheric layers. 

\section{SIMULATIONS}

\subsection{Initial Model and the Set of Simulations} 

The initial axisymmetric model for all the calculations is the same as 
that used by \citet{mejia05}. The central star is 0.5 $M_\odot$, and 
the nearly Keplerian disk of 0.07 $M_\odot$ has a surface density 
$\Sigma(r) \propto r^{-0.5}$ from 2.3 AU to 40 AU. The initial grid 
has (256, 128, 32) cells in ($r,\phi,z$) above the midplane. 
When the disk expands at the onset of GI's, the grid is 
extended radially and vertically. The initial minimum value of the 
Toomre stability parameter $Q$ is about 1.5, and so the disk is 
marginally unstable to GI's.  The initial model is seeded with 
low-amplitude random density noise. We use $T_{irr} = 15$ K, which
is lower than the 50 K assumed in \citet{boss02} because our larger and 
less massive disk is mostly stabilized by $T_{irr} = 50$ K.
In this paper, we present simulations with four metallicities Z = 1/4 Z$_\odot$ 
(one-quarter solar metallicity), 1/2 Z$_\odot$, Z$_\odot$, and 2 Z$_\odot$.
The 1/4 Z$_\odot$ simulation was started from the 1/2 Z$_\odot$ disk 
after 13.0 outer rotation periods (ORPs) of evolution, to save computing 
resources. Here 1 ORP (about 250 yrs) is the initial rotation period at 33 AU. 
The varied metallicity cases use a maximum grain size 
$a_{max} = 1 {\mu}$m in the dust opacities.   
An additional simulation with $a_{max} = 1$ mm and Z = Z$_{\odot}$ 
is conducted to explore the effects of grain growth. 

\subsection{Results}

The current calculations resemble those presented in \citet{mejiaphd} 
and \citet{mejia05}.  The disks remain fairly axisymmetric until a 
{\sl burst phase} of rapid growth in nonaxisymmetric structure. 
Subsequently, the disks gradually transition into a quasi-steady {\sl 
asymptotic phase}, 
where heating and cooling are in rough balance, and average 
quantities change slowly \citep[see also][]{lodato05}. 
Table \ref{comparetable} summarizes some of the results.  In the
table, Duration refers to the simulation length measured in ORPs, 
$t_1$ is the time in ORPs at which the burst phase begins, 
$t_2$ is the approximate time in ORPs when the simulation enters 
the asymptotic state, $\langle A \rangle$ is a time-averaged integrated 
Fourier amplitude for all nonaxisymmetric structure (see below), 
$t_{cool}$ is the final global cooling time obtained by dividing the 
final total internal energy of the disk by the final total luminosity, 
and Thin$\%$ is the percentage of disk volume that is optically thin 
during the asymptotic phase.

One noticeable effect is that the onset
of the burst phase ($t_1$) is delayed for higher metallicity and larger 
grain size (Table \ref{comparetable}), as expected due to higher opacity 
and therefore slower cooling. Note that, over the bulk of our large 
cool disk, increasing $a_{max}$ increases the opacity. Although the time
to reach the asymptotic phase is relatively insensitive to grain size
and metallicity, the overall final $t_{cool}$ listed
in Table \ref{comparetable} illustrates that the correlation between 
cooling time and opacity carries over to late times. 
During the asymptotic phase, in all cases, the Toomre $Q$ values 
remain roughly constant with time, with values ranging between 
1.3 to 1.8 for $r =$ 10 to 40 AU, and the mass inflow rates peak near
15 AU at  $\sim 10^{-6} {M_\odot}$/yr, with negligible difference
between 1/2 Z$_{\odot}$ and 2 Z$_{\odot}$ to the accuracy 
that we can measure these inflows \citep{mejia05}. 
Although there are some regions of superadiabatic gradients, the 
rapid overall convective cooling reported by \citet{boss02,boss04} does 
not occur. We do see upward and downward motions, which we 
attribute to hydraulic jumps \citep{boley05}. Whether or not some of 
these motions are actually thermal convection, they do not result in 
rapid cooling for our disks.

\begin{figure}

\plotone{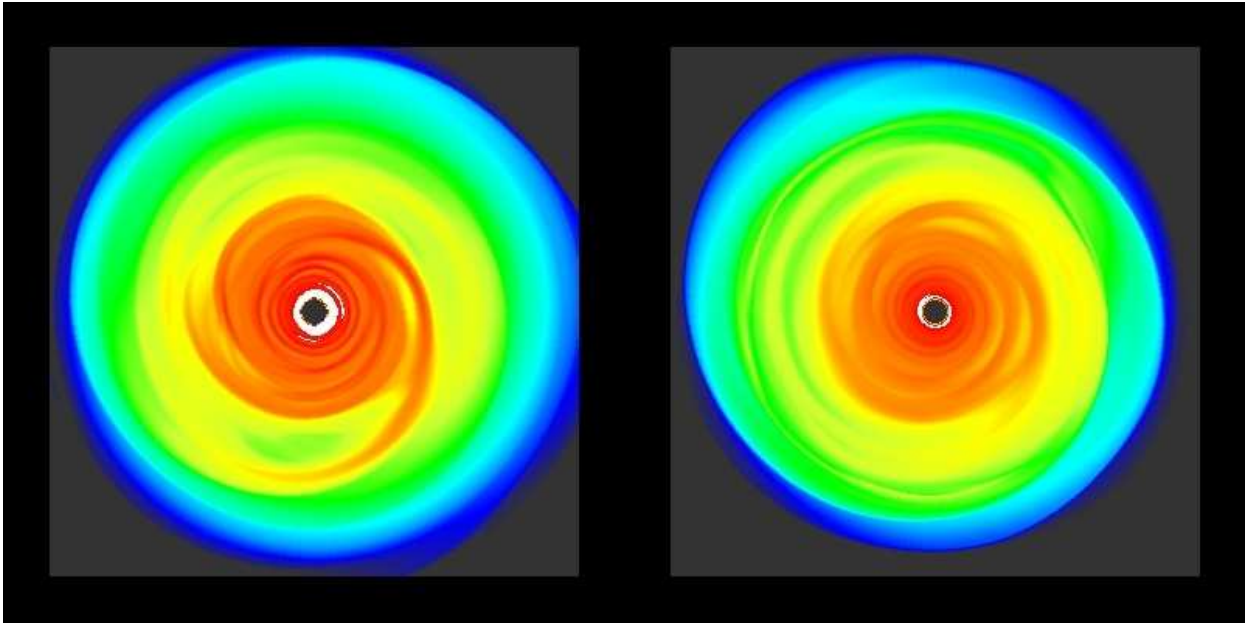}

\caption{Midplane density maps at 15 ORPs for the 1/4 Z$_{\odot}$
(left panel) and 2 Z$_{\odot}$ (right panel) simulations. 
Each square is 113 AU on a side. 
Densities are displayed on a logarithmic scale running from light grey
to black (print version) or dark blue to dark red (online version), 
as densities range from about 4.8$\times$10$^{-16}$ to 
4.8$\times$10$^{-11}$ g cm$^{-3}$, respectively,
except that both scales saturate to white at even higher densities.
\label{den} }
\end{figure}

In Figure \ref{den}, which shows midplane densities at 15 ORPs, 
the spiral structure appears stronger  
for 1/4 Z$_\odot$ than for 2 Z$_\odot$. In order to quantify 
differences in GI strength, we compute integrated Fourier amplitudes 
\citep{imamura00}

\begin{displaymath}
 A_m = \frac{{\int \rho_m rdrdz}}{{\int \rho_0 rdrdz}}, 
\end{displaymath}

\noindent
where $\rho_0$ is the axisymmetric component of the density and 
$\rho_m$ is the amplitude of the cos($m\phi$) component. 
Although variable, after $\sim$ 14 ORPs, the $A_m$'s for 
most $m$'s are greater for 1/4 Z$_\odot$ than for higher Z's. 
To measure total nonaxisymmetry, we sum the $A_m$'s and 
average this sum over 14.5 to 15.5 ORPs.  As shown in 
Table \ref{comparetable}, this global measure $\langle A \rangle$ is 
greatest for
1/4 Z$_\odot$ and generally decreases with increasing metallicity
and grain size. 

\begin{figure}
\epsscale{.80}
\plotone{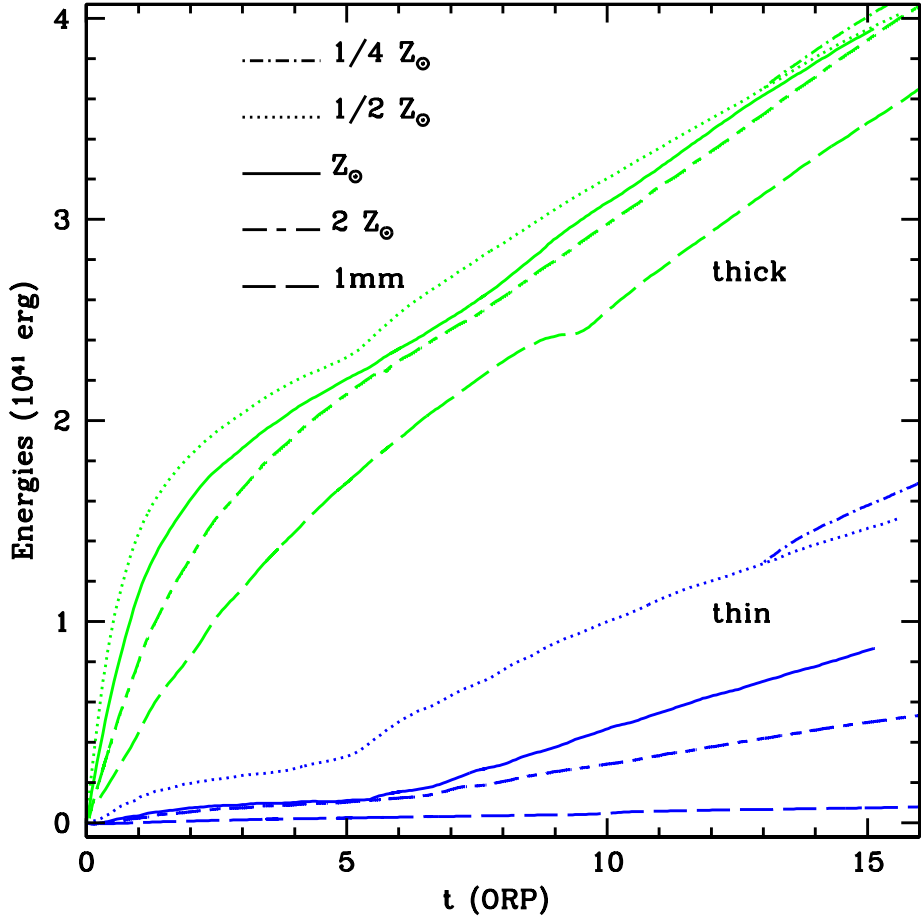}
\caption{Cumulative total energy loss as a function of time due to 
radiative cooling in optically thick (upper set, labelled ``thick'') and 
optically thin regions (lower set, labelled ``thin''). Both of these are {\sl net} 
global cooling after heating by irradiation is subtracted. The curves labeled
by a metallicity value all use $a_{max} =$ 1 $\mu$m. The curves 
labeled ``1mm'' are for a calculation with $a_{max} =$ 1mm and 
solar metallicity. Note that 
the 1/4 Z$_{\odot}$ run starts from the 1/2 Z$_{\odot}$ simulation at 
about 13 ORPs. A color version of this figure appears on line.  
\label{energy}}
\end{figure}

Figure \ref{energy} plots the cumulative energy loss 
due to cooling computed for only half the disk 
as a function of time.  The upper curves show energy loss 
from the disk interior after compensating for energy 
input by residual irradiation and by the glowing disk upper atmosphere; 
the lower curves show net energy loss from optically thin regions after 
accounting for heating due to envelope irradiation. Due to our restricted
vertical resolution and use of the Eddington atmosphere fit over one or two 
vertical cells, the ``thick'' curves effectively include most of the 
photospheric layers for most columns through the disk. The ``thin'' curves tally additional cooling from extended layers above the photospheric cells, usually 
with $\tau_R << 1$, plus the parts of the outer disk that are optically thin 
all the way to the midplane. 
The initial cooling rates for the optically thick regions plus 
photosphere clearly differ. In fact, the initial slopes 
of the ``thick'' curves give $t_{cool} \sim $ Z/Z$_{\odot}$ ORPs. 
However, the initial disks are far from radiatively relaxed, and so there 
are transients. Remarkably, by the asymptotic phase, all the 
disk interior-plus-photosphere curves converge to similar energy loss rates.  
During these late times, the differences between the total cooling rates are 
dominated by the optically thin regions, which are larger for the lower 
metallicity cases, as indicated by the Thin$\%$ entry in 
Table \ref{comparetable}. The overall asymptotic phase $t_{cool}$'s in 
Table \ref{comparetable}, based on summing the thick and thin 
loss rates, are longer for higher metallicity and larger grain size.
Altogether, the evidence in Table \ref{comparetable} and 
Figures \ref{den} and \ref{energy} shows that higher opacity leads to 
slower cooling and that slower cooling produces lower GI amplitudes.
We remind the reader that we detect these differences over a 
much narrower range of metallicities (1/4 to 2 Z$_{\odot}$) than 
considered by \citet{boss02} (0.1 to 10 Z$_{\odot}$). 

As in \citet{mejiaphd}, except for brief transients during the burst 
phases of some runs, these disks do not form dense clumps, in apparent 
disagreement with \citet{boss02}. To investigate whether the disk evolution 
depends on spatial resolution in the asymptotic phase \citep{boss00, boss05, 
pickett03}, both the 1/4 Z$_{\odot}$ and 2 Z$_{\odot}$ simulations
are extended for another 2 ORP's with quadrupled azimuthal resolution 
(512 zones), and the disks do not fragment into dense clumps. This is consistent 
with the analytic arguments in \citet{rafikov05} that an unstable disk and fast 
radiative cooling are incompatible constraints for realistic disks at 10 AU  
(see Boss 2005 for a different perspective). Indeed, if $t_{cool}$ listed 
in Table~\ref{comparetable} is a good measure of local cooling times in 
these disks, we do not expect fragmentation. \citet{gammie01} 
shows that fragmentation occurs only if the local $t_{cool}$ is less 
than about half the local disk orbit period $P_{rot}$  
\citep[see also][]{rice03, mejia05}, except possibly near sharp opacity 
edges \citep{johnson03}. We only find locallized cooling times 
shorter than 0.5 $P_{rot}$ in the asymptotic phase of the 2 Z$_{\odot}$ case, 
and then only in the 30 to 40 AU region, which is optically thin. This occurs 
because, even though $t_{cool} \sim$ Z in optically thick regions (higher optical 
depth), $t_{cool} \sim$ Z$^{-1}$ in thin ones (more emitters). As a result, this 
disk displays the steepest drop of local $t_{cool}$ with $r$. The short 
local $t_{cool}$'s appear to be highly variable and transient. The continuation 
of this simulation for 2 ORPs at higher azimuthal resolution (512 zones) 
does not show evidence for fragmentation into clumps. It could prove 
important to push our simulations to higher Z in the future.

\section{DISCUSSION}

Our results show that GI strength decreases as 
metallicity increases and, contrary to \citet{boss02}, 
that global radiative cooling is too slow for 
fragmentation into dense clumps. In the asymptotic phase, cooling rates 
for the disk interior plus photospheric layers converge for all Z, but the 
total cooling, including the optically thin regions, is higher for lower Z. 
Thus, the optically thin upper atmosphere and outer disk play a role in the 
cooling rate of an evolved disk.  In fact, the fractional volume of the 
optically thin regions becomes very large at late times 
(see Thin$\%$ in Table \ref{comparetable}). Note also that the optically 
thick region fractional volume,  $1 - {\rm Thin}\%$,  varies roughly as Z.  
The greater surface area of the disk photosphere at higher Z tends 
to compensate for the higher opacity and leads to convergence of the
cooling rates for the parts of the disk contained within the 
photospheric layers. In this respect, we confirm Boss's conclusion
that the outcome of the radiative evolution is somewhat insensitive to 
metallicity. However, the important difference is that we do not see
fragmentation into dense clumps, presumably because our cooling 
rates are much lower than in \citet{boss02}. 
For the 1mm case, the optically thin regions have a much 
smaller volume (Table 1) and contribute little to cooling. 
Outside the inner few AU, bigger grains make the disk more opaque to 
longer wavelengths, and $t_{cool}$ is thus considerably larger, even 
initially. 

Our results argue against direct formation of gas giants by 
disk instability in two ways -- the global radiative cooling times seem too
long for fragmentation to occur and GI's are stronger overall for {\sl lower}
metallicity. Nevertheless, it is still possible that GI's play 
an important role in gas giant planet formation.
\citet{durisen05} suggest that dense gas rings produced by GI's will 
enhance the growth rate of solid cores by drawing solids toward 
their centers \citep{haghighipour03} and thereby accelerating core
accretion. Such rings are indeed produced 
in the inner disks of all our calculations regardless of metallicity or grain 
size, and they appear to be still growing when the calculations end.  
In the weaker GI environments of high metallicity, there is less
self-gravitating turbulence to interfere with the radial drift of solids 
\citep{durisen05}. In this way, rings may provide a natural shelter and 
gathering place for growing embryos and cores. 

The apparent disagreement between our results 
and those of \citet{boss02, boss04} could be due to
any number of differences in techniques and assumptions, such as
artificial viscosity, opacities, equations of state, initial disk models and
perturbations, grid shapes and resolution, and radiative boundary
conditions, including the way that we handle irradiation. We are
now collaborating with Boss in an effort to pinpoint which of these is
the principal cause (K. Cai et al., in preparation). Preliminary results 
suggest that it is the radiative boundary conditions. We are therefore
developing alternative techniques for disk radiative transfer  
that we hope are more reliable and accurate.

\acknowledgments
We thank A.P. Boss and an anonymous referee for useful comments.
This work was supported in part by NASA Origins of Solar Systems grants 
Nos. NAG5-11964 and NNG05GN11G, 
by NASA Planetary Geology and Geophysics grant 
No. NAG5-10262, and by a Shared University Research grant from 
IBM, Inc. to Indiana University.

\clearpage

\begin{table}[H]
\begin{center}
\caption{\label{comparetable} Simulation Results}
\vspace{0.3cm}
\begin{tabular}{|c|c|c|c|c|c|c|c|c|}
\tableline\tableline
Case & $f_\kappa$ & $a_{max}$ & 
Duration\tablenotemark{a} & $t_1$\tablenotemark{a} &
$t_2$\tablenotemark{a} & $\langle A \rangle$ & $t_{cool}$ 
\tablenotemark{a} & Thin\% 
\\
\tableline
1/4 Z$_\odot$ & 1/4 & 1 ${\mu}$m   & 3.8\tablenotemark{b}  & N/A & N/A & 
1.29  & 2.1 & 99\% \\
1/2 Z$_\odot$ & 1/2 & 1 ${\mu}$m   & 15.6 & 4.0  & 10 & 1.09  & 2.9 & 
98\%\\
Z$_\odot$ & 1.0 & 1 ${\mu}$m & 15.7 & 5.0 & 10 & 1.10 & 3.2 & 94\%\\
2 Z$_\odot$ & 2.0 & 1 ${\mu}$m     & 16.5 & 5.0 & 10 & 0.72 & 3.7 & 
86\%\\
1mm & 1.0 & 1 mm         & 17.2 & 7.0 & 11 & 0.88 & 4.5 & 44\%\\
\tableline
\end{tabular}
\tablenotetext{a}{All times are given in units of ORPs.}
\tablenotetext{b}{From 13.0-16.8 (ORPs). }
\end{center}
\end{table}

\clearpage

\clearpage

\end{document}